\documentclass[12pt,a4paper]{scrartcl}
\pdfoutput=1
\usepackage[utf8]{inputenc}
\usepackage[sorting=none,%
  citestyle=numeric-comp]{biblatex}
\usepackage[T1]{fontenc}
\usepackage{amsmath,amssymb,mathtools}
\usepackage[separate-uncertainty=true,retain-unity-mantissa=false]{siunitx}
\usepackage{slashed}
\usepackage{graphicx}
\usepackage{booktabs,multicol,multirow}
\usepackage[shortlabels]{enumitem}
\usepackage{hyperref}
\usepackage{xspace}
\usepackage[noblocks]{authblk}
\usepackage[dvipsnames]{xcolor}
\usepackage{soul}
\usepackage[textsize=tiny]{todonotes}
\usepackage{subcaption}
\usepackage{ulem}
\usepackage{bbold}
\usepackage{comment}
\usepackage[compat=1.0.0]{tikz-feynman}

\newcommand{\shad}{\sigma_{\mathrm{had}}}

\addtokomafont{disposition}{\rmfamily}

\title{\Large Resolving the muon $g-2$ tension through $Z'$-induced modifications to $\sigma_{\mathrm{had}}$}
\author[1,4]{Nina M. Coyle\footnote{\href{mailto:ncoyle@uci.edu}{ncoyle@uci.edu}}}
\author[1,2,3]{Carlos E.M. Wagner\footnote{\href{mailto:cwagner@uchicago.edu}{cwagner@uchicago.edu}}}
\affil[1]{University of Chicago and Enrico Fermi Institute, 5720 South Ellis Avenue, Chicago, IL~60637~USA}
\affil[2]{Kavli Institute for Cosmological Physics, University of Chicago, Chicago, IL 60637}
\affil[3]{HEP Division, Argonne National Laboratory, 9700 Cass Ave., Argonne, IL 60439, USA}
\affil[4]{Department of Physics and Astronomy, University of California, Irvine, CA 92697}
\date{}
\titlehead{\hfill \parbox{3cm}{\vspace{-2cm}\begin{flushright} 
\texttt{EFI 23-3} \end{flushright}}}

\begin{document}
\maketitle

\begin{abstract}\noindent
The QED hadronic vacuum polarization function plays an important role in the determination
of precision electroweak observables and of the anomalous magnetic moment of the muon. 
These contributions have been computed from data, by means of dispersion relations affecting
the electron positron hadronic cross sections, or by first principle lattice-QCD computations
in the Standard Model.  Today there is a discrepancy between the two approaches for determining these contributions, which affects the comparison of the measurement of the anomalous
magnetic moment of the muon with the theoretical predictions. In this article, we revisit
the idea that this discrepancy may be explained by the presence of a new light gauge boson
that couples to the first generation quark and leptons and has a mass below the GeV scale. We discuss the  requirements for its consistency with observations and the phenomenological implications of such a gauge extension. 
   \end{abstract}
\setcounter{footnote}{0}

\newpage

\tableofcontents

\newpage


\section{Introduction}
\label{sec:intro}

The Standard Model (SM) provides a very good description of all elementary particle interactions observed at laboratory experiments.  It is therefore important to test this model by precision measurements of the interactions of quarks and leptons with the Higgs and SM gauge fields. In particular,   
the measurement of the muon anomalous magnetic moment by the Muon $g-2$ collaboration at Fermilab~\cite{Muong-2:2021ojo} , together with the previous Brookhaven measurement of this quantity~\cite{Muong-2:2004fok}, shows a 4.2~$\sigma$ deviation with respect to the SM predictions~\cite{Aoyama:2020ynm} and provides an intriguing hint towards new physics. An updated measurement from the Muon $g-2$ collaboration~\cite{Muong-2:2023cdq} further reinforces the discrepancy.
Reconciling the theoretical prediction for $a_{\mu} = \frac{g_\mu-2}{2}$ with the experimental results has recently been a productive topic of examination. Many investigations have proposed models with direct modifications that enhance the muon magnetic moment (see e.g.~\cite{Moroi:1995yh,Greljo:2021xmg,Yang:2022gvz,Martin:2001st,Liu:2018xkx,Baum:2021qzx,Chakraborti:2021kkr,Athron:2021iuf,Babu:2022pdn,Arcadi:2021cwg,Wang:2021bcx}).

However,  there is disagreement between the theoretical predictions for $a_{\mu}$  obtained by computing the hadronic vacuum polarization contribution through two different methods: first through dispersion relations of the experimental observations of the cross section $\shad = \sigma(e^+ e^- \rightarrow {\mathrm{hadrons}})$, and second through the lattice-QCD prediction computed by the BMW collaboration~\cite{Borsanyi:2020mff}. While the value of $a_{\mu}$ derived from the hadronic cross section observations, which we will denote as $a_{\mu}^{ee}$, disagrees with the $g-2$ and Brookhaven combined measurement $a_{\mu}^{\mathrm{exp}}$ at the previously mentioned 4.2~$\sigma$ level, the anomalous magnetic moment derived from the BMW lattice-QCD calculation is consistent with $a_{\mu}^{\mathrm{exp}}$ at the 1.6~$\sigma$ level. Although the total result for $a_{\mu}^{\mathrm{HVP}}$ obtained by the BMW collaboration has yet to be confirmed, the contribution from the intermediate time window, which accounts for about a third of $a_{\mu}^{\mathrm{HVP}}$, has been successfully tested by several lattice-QCD groups~\cite{Aubin:2022hgm,Ce:2022kxy,
Alexandrou:2022amy,Wang:2022lkq,Blum:2023qou,Bazavov:2023has}.

It is therefore conceivable that one may partially resolve the $g_\mu-2$ anomaly by introducing new physics that reconciles the hadronic vacuum polarization effects extracted from $e^+ e^-$ observations with the computation by the BMW collaboration. Previous works have examined this possibility~\cite{Keshavarzi:2018mgv,Crivellin:2020zul,DiLuzio:2021uty}, concluding that  the modification of the hadronic vacuum polarization by new physics is ruled out by electroweak precision measurements or by the experimental constraints on the hadronic and leptonic couplings of the required light gauge boson. An alternative approach introducing a new Z' coupled to electrons and muons and with a mass very close to  the KLOE center of mass energy, leading to not only a change in $g_\mu-2$ by new physics but also to a modification of the KLOE luminosity determination, has also been proposed to reconcile the theoretical and experimental determinations of $g_\mu-2$~\cite{Darme:2021huc,Darme:2022yal}. Additional long-lived neutral hadrons have also been proposed as a solution to this discrepancy~\cite{Farrar:2022vrs}. 

In this work, we focus on the approach proposed by~\cite{DiLuzio:2021uty}, which introduces a $Z'$ boson coupled to first-generation leptons and quarks. While the original work concluded that LEP-2, BaBar, and isospin breaking observables constrain the model such that an explanation of $g-2$ is impossible, we note a few modifications and caveats that may avoid these bounds, and discuss further challenges to make this framework a realistic one.

This paper is structured as follows. In Section \ref{sec:background}, we provide more detailed background on the $g_{\mu}-2$ anomaly and the disagreement in $\sigma_{\mathrm{had}}$, and introduce the $Z'$ model. In Section \ref{sec:constraints}, we address the relevant constraints on this model and the manner in which they may be avoided. We further examine the consequences for $g_\mu-2$ in Section~\ref{sec:muong2}. We reserve Section \ref{sec:conclusions} for our conclusions. 


\section{Background and model}
\label{sec:background}

\subsection{The $g_\mu-2$ anomaly}
The anomalous magnetic moment of the muon is defined as $a_{\mu} = (g_\mu-2)/2$, and has been measured by the E821 and Fermilab Muon $g-2$ experiments to be~\cite{Muong-2:2021ojo,Muong-2:2004fok}
\begin{align}
    a_{\mu}^{\mathrm{exp}} = 116 \; 592 \; 061(41) \times 10^{-11}.
\end{align}
This result is in tension with the SM estimate from the Muon $g-2$ theory initiative ~\cite{Aoyama:2020ynm},
\begin{align}
    a_{\mu}^{\mathrm{SM}} = 116 \; 591 \; 810(43) \times 10^{-11} .
\end{align}
The SM contributions to $a_{\mu}$ may be broken down as
\begin{align}
    a_{\mu}^{\mathrm{SM}} = a_{\mu}^{\mathrm{QED}} + a_{\mu}^{\mathrm{EW}} + a_{\mu}^{\mathrm{LbL}} + a_{\mu}^{\mathrm{HVP}},
\end{align}
where the superscript $\mathrm{QED}$, $\mathrm{EW}$, $\mathrm{LbL}$ and $\mathrm{HVP}$ refer to the pure QED, electroweak Higgs and gauge boson, hadronic light by light, and the hadronic vacuum polarization contributions, respectively. 
Of these contributions, the least well understood is the one coming from the hadronic vacuum polarization effects,  $a_{\mu}^{\mathrm{HVP}}$, which is determined using a compilation of experimental data on $\shad$. In this analysis, the contribution from loop integrals containing insertions of HVP in the photon propagator can be extracted from dispersion integrals over the cross section of a virtual photon decaying to hadrons. In particular, the calculation in terms of $\shad$ is
\begin{align}
    (a_{\mu}^{\mathrm{HVP}})_{ee} = \frac{1}{4\pi^3} \int_{m^2_{\pi^0}}^{\infty} ds \ K(s) \ \shad (s),
\end{align}
with a kernel $K(s) \simeq m_\mu^2/(3 s)$ for $s \gg m_\mu^2$.
Based on the observed $\shad$, one finds a hadronic vacuum polarization contribution of
\begin{align}
    (a_{\mu}^{\mathrm{HVP}})_{ee} = 6931(40)\times 10^{-11}.
\end{align}
However, a recent lattice-QCD calculation by the BMW collaboration finds the hadronic vacuum polarization contribution to be~\cite{Borsanyi:2020mff}
\begin{align}
    (a_{\mu}^{\mathrm{HVP}})_{\mathrm{BMW}} = 7075(55)\times 10^{-11} .
\end{align}

Thus, in addition to the tension between the experimentally-measured muon anomalous magnetic moment and the SM prediction, there is also a tension between theory predictions. In fact, while the measured value of $a_{\mu}$ and the SM prediction derived from $\shad$ are in tension at more than the 4.2~$\sigma$ level, there is disagreement at only 1.6~$\sigma$ between the experimental measurement and the BMW prediction. As such, one may be able to resolve the $g_{\mu}-2$ tension by modifying $\shad$ rather than directly enhancing $a_{\mu}$. Recent results from the CMD-3 experiment \cite{CMD-3:2023alj} have found a larger HVP contribution to $a_{\mu}$ from the $e^+e^- \to \pi^+\pi^-$ cross section, potentially lessening the tension between the two calculations. A careful analysis that accounts for the various experimental approaches is required to properly combine these results; in this work we focus on resolving the discrepancy between the previous result from the $g-2$ theory initiative, specifically employing data from CMD-2 and KLOE, and the BMW lattice-QCD calculation as an illustration. To reconcile these calculations, one may introduce new physics that interferes with the SM contribution to $\shad$ in order to suppress the observed rate. 

In particular, we wish to source a shift in $\shad$ such that one obtains
\begin{align}
 \Delta a_{\mu}^{\mathrm{HVP}} = (a_{\mu}^{\mathrm{HVP}})_{\mathrm{BMW}} -
(a_{\mu}^{\mathrm{HVP}})_{ee} =
1.44(68) \times 10^{-9},
\end{align}
where we have combined the uncertainties in quadrature. This implies that the observed hadronic cross section is  smaller than the one expected by the BMW lattice-QCD hadronic vacuum polarization determination.  
A modification to the hadronic cross section from new physics, denoted by $\Delta \sigma_{\mathrm{had}}$, will need to be subtracted off from the computation of the HVP contribution to the muon magnetic moment, giving a shift of
\begin{align}\label{eq:damu}
\Delta a_{\mu}^{\mathrm{HVP}} = \frac{1}{4\pi^3}\int_{m^2_{\pi^0}}^{\infty} ds \; K(s) (-\Delta \shad (s)).
\end{align}
We will analyze the new physics which may give a destructive interference to the observed $\shad$, thus sourcing a positive shift in the expected $a_{\mu}^{\mathrm{HVP}}$. 

\subsection{Model and parameter values}
In this paper, we focus on the addition of a $Z'$ with vector couplings to the electron and first generation quarks, as proposed in Ref.~\cite{DiLuzio:2021uty}. We start with the interactions:
\begin{align} \label{eq:Lagrangian}
    \mathcal{L} \supset (g_V^e \bar{e} \gamma^{\mu} e + g_V^u \bar{u} \gamma^{\mu} u + g_V^d \bar{d} \gamma^{\mu} d)  Z'_{\mu},
\end{align}
which may be understood as an effective theory, without any assumed structure in the values of $g_V^{e,u,d}$.
Given the behavior of the kernel $K(s)$, which goes approximately as $1/s$, the greatest contribution to $a_{\mu}$ from $\shad$ comes from the $\sqrt{s} \lesssim 1$ GeV region, in which the dominant process is $e^{+} e^{-} \rightarrow \pi^+ \pi^-$. 
This region of energies is also preferred by precision electroweak measurements, 
since corrections to the hadronic cross section at higher energies lead to sizable 
corrections to $\alpha(M_Z)$, the electromagnetic structure constant at the $M_Z$ scale~\cite{Keshavarzi:2020bfy,Crivellin:2020zul}.

\begin{figure}[t] 
\centering
\includegraphics[width=0.6\textwidth]{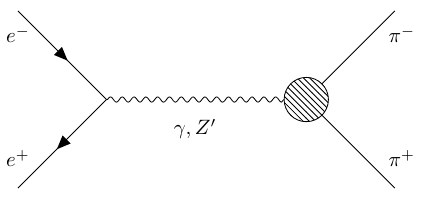}
\caption{Relevant Feynman diagram for $\gamma$- and $Z'$-mediated $e^+e^- \to \pi^+ \pi^-$ scattering.}
\label{fig:ee_to_pipi}
\end{figure}

As such, the primary $Z'$ process of interest for $\sigma_{\mathrm{had}}$ is the process $e^{+} e^{-} \rightarrow Z' \rightarrow \pi^+ \pi^-$. Given the above Lagrangian  and taking the $Z'$ to mix with the vector mesons in a manner similar to the photon, the modification from an off-shell $Z'$ to the SM cross section for electron-positron annihilation to pions may be written as~\cite{DiLuzio:2021uty}
\begin{align} \label{eq:ee_sig_ratio}
\frac{\sigma^{SM+Z'}_{\pi\pi}}{\sigma_{\pi\pi}^{\mathrm{SM}}} (s) = \left| 1 - \frac{g_V^e(g_V^u - g_V^d)}{e^2} \frac{s}{s-m_{Z'}^2 + i m_{Z'} \Gamma_{Z'}} \right|^2.
\end{align}
We show the relevant diagram in Fig~\ref{fig:ee_to_pipi}; the mixing with the $\rho$, which couples to the $\pi^+\pi^-$ final state, is included in the effective vertex defining the pion form factor and will be discussed in more detail in later sections.

The width $\Gamma_{Z'}$ has two main contributing decays: to $\pi^+ \pi^-$ and $e^+ e^-$. The decay widths for the electron and pion decays are given by
\begin{subequations}\label{eq:widths}
    \begin{align}
    \Gamma(Z' \to e^+ e^-) &= \frac{1}{3} \frac{(g_V^e)^2}{4\pi} m_{Z'} \sqrt{1-4\frac{m_e^2}{m_{Z'}^2}}\left(1+2\frac{m_e^2}{m_{Z'}^2}\right) \\    
    \Gamma(Z' \to \pi^+ \pi^-) &= \frac{1}{3} \frac{(g_V^{ud})^2}{4\pi} m_{Z'} \sqrt{1-4\frac{m_{\mu}^2}{m_{Z'}^2}}\left(1+2\frac{m_{\mu}^2}{m_{Z'}^2}\right) R(m_{Z'}^2),
    \end{align}
\end{subequations}
where for convenience we have defined $g_V^{ud} \equiv g_V^u - g_V^d$, and $R(s)$ is defined as
\begin{align} \label{eq:R(s)}
R(s) = \frac{\sigma^{\mathrm{SM}}_{e^+e^- \to \mathrm{had}}}{\sigma^{\mathrm{SM}}_{e^+e^- \to \mu^+ \mu^-}}(s),
\end{align}
where $\sigma^{\mathrm{SM}}_{e^+e^- \to \mathrm{had}}$ is the SM cross section to hadrons; because the cross section to hadrons is dominated by the $e^+e^- \to \pi^+\pi^-$ process within the range of energies we are interested in, we will refer to this quantity as $\sigma_{\pi\pi}^{\mathrm{SM}}$. We emphasize that in the absence of any additional decay channels, the width $\Gamma_{Z'}$ must be calculated from the couplings using Eq.~(\ref{eq:widths}) and depends on the specific choices for $g_V^e$ and $g_V^{ud}$. \\

Let us define the effective coupling $\tilde{g} \equiv -g_V^e(g_V^u-g_V^d)/e^2$, and denote the observed cross section by $\sigma_{\pi\pi}^{\mathrm{obs}}$. The resulting $\sigma_{\pi\pi}^{\mathrm{obs}}$ is given by:
\begin{align}
    \sigma_{\pi\pi}^{\mathrm{obs}}(s) = \sigma_{\pi\pi}^{\mathrm{SM}}(s) \left( 1 + \frac{\tilde{g}^2 s^2 + 2\tilde{g} s (s - m_{Z'}^2)}{(s-m_{Z'}^2)^2 + m_{Z'}^2 \Gamma_{Z'}^2} \right) \equiv \sigma_{\pi\pi}^{\mathrm{SM}}(s) \; \big(1+\delta(\tilde{g},s)\big)
\end{align}
and thus the $Z'$ modification to the observed cross section is given by
\begin{align}
\Delta \sigma_{\pi\pi}^{\mathrm{NP}}(s) &= \sigma_{\pi\pi}^{\mathrm{obs}}(s) - \sigma_{\pi\pi}^{\mathrm{SM}}(s) \nonumber \\
&= \sigma_{\pi\pi}^{\mathrm{obs}}(s) \left(\frac{\delta(\tilde{g},s)}{1+\delta(\tilde{g},s)} \right)
\end{align}
We can immediately see that the new physics contribution can interfere destructively with the SM depending on the relative signs of $\tilde{g}$ and $s-m_{Z'}^2$.\\

We interpolate the CMD-2~\cite{Aulchenko:2006dxz,CMD-2:2006gxt} and SND~\cite{Achasov:2006vp} $\sigma(e^+ e^- \to \pi^+\pi^-)$ data and the KLOE~\cite{KLOE-2:2017fda} and BaBar~\cite{BaBar:2009wpw} $\sigma(e^+e^- \to \pi^+ \pi^- (\gamma))$ results to obtain a curve for each of the observed $\sigma_{\pi\pi}^{\mathrm{obs}}$ employed by the $g-2$ theory initiative. These data sets represent two different types of experiments: CMD-2 and SND are energy scan experiments, while KLOE and BaBar use initial-state radiation (ISR) to gather data for a range of energies. These two experimental approaches have complementary advantages and disadvantages. While energy scan experiments have high energy resolution, they operate at fixed center of mass energies and therefore lack data for energy ranges between data points; ISR experiments, on the other hand, have lower resolution due to binning, but yield a continuous measurement of the cross section. We examine how a $Z'$ would affect each type of experiment.

Although the absence of a visible resonant feature
similar to a $Z'$ peak in the cross section data 
may be due to the low experimental resolution in certain energy regimes of these experiments, in our work we will assume that these features may be observed and therefore demand the absence of any unexplained compensating 
feature in the SM hadronic cross section. One manner in which this may be achieved is by an enhancement of the width by taking
larger $g_V^{ud}$ and a $Z'$ mass near the $\rho$ resonance mass of 770 MeV. As we will see in Section~\ref{sec:constraints}, the pion mass difference places constraints on the values of $g_V^{ud}$. Although a full non-perturbative analysis is in order to determine the precise bounds, an estimate based on the Cottingham method~\cite{Cottingham:1963zz,Donoghue:1996zn,Stamen:2022uqh,Crivellin:2022gfu} leads to allowed values of order $g_{V}^{ud} \lesssim 0.1$. We thus find benchmark points which obtain $\Delta a_{\mu}^{\mathrm{HVP}}=1.44\times10^{-9}$ for $m_{Z'} = 0.79$ GeV and $g_V^{ud}=0.1$ of
\begin{align}\label{eq:unfixed_gud_benchmarks}
    g_V^e =
    \begin{cases}
        -3.3 \times 10^{-3} \;\;\; \mathrm{(CMD-2)} \\
        -2.6 \times 10^{-3} \;\;\; \mathrm{(KLOE)} \\
        -3.4 \times 10^{-3} \;\;\; \mathrm{(SND)} \\
        -2.7 \times 10^{-3} \;\;\; \mathrm{(BaBar)}
    \end{cases}
\end{align}

\begin{figure}
    \centering
    \includegraphics[width=0.9\textwidth]{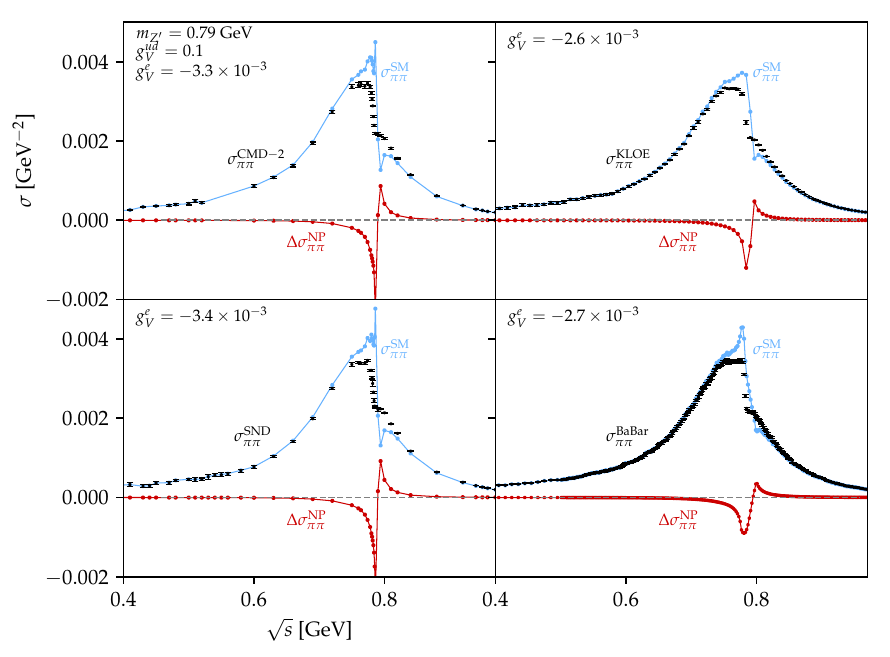}
    \caption{New physics modifications to CMD-2 (upper left), KLOE (upper right), SND (lower left), and BaBar (lower right) which give $\Delta a_{\mu}^{\mathrm{HVP}}=1.44\times10^{-9}$ with $m_{Z'}=0.79$ GeV and $g_V^{ud}=0.1$. The values of the electron coupling are $g_V^e = \{-3.3,-2.6,-3.4,-2.7\} \times 10^{-3}$ for CMD-2, KLOE, SND, and BaBar, respectively. The red lines show the linear interpolation of the $\Delta \sigma^{\mathrm{NP}}_{\pi\pi}$ points, which is used to estimate the resulting $\Delta a_{\mu}^{\mathrm{HVP}}$. We also show the resulting inferred SM cross section (blue).}
    \label{fig:dsigma_low_gud}
\end{figure}

\begin{figure}
    \centering
    \includegraphics[width=0.55\textwidth]{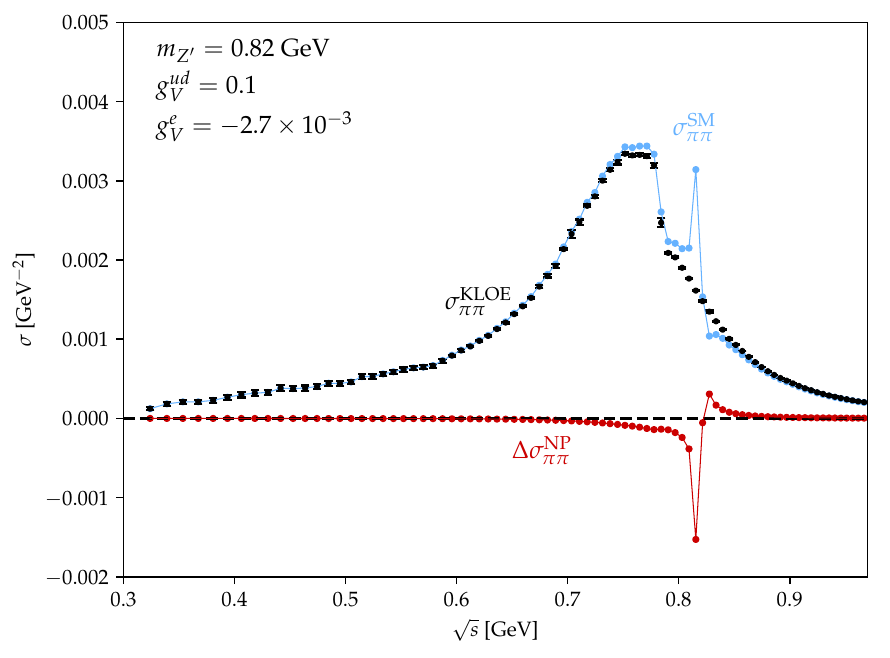}
    \caption{New physics modifications to KLOE for $m_{Z'}=0.82$ GeV with $g_V^{e}=-2.7\times10^{-3}$ and $g_V^{ud}=0.1$.}
    \label{fig:dsigma_KLOE_82}
\end{figure}

In calculating the impact of $\Delta \sigma^{\mathrm{NP}}_{\pi\pi}$ on the observed data and the resulting inferred SM cross section, we account for a resolution of $10^{-3}\sqrt{s}$ in the CMD-2 and SND experiments, a resolution of 0.002 GeV$^2$ and a binning of 0.01 GeV$^2$ in $s$ for the KLOE data, and a resolution of 0.006 GeV and a binning of 0.002 GeV in $\sqrt{s}$ for the BaBar data. To account for the energy resolution of each experiment, we smear the $\Delta \sigma$ curve in $s$ using a Gaussian with a width determined by the respective experimental resolution; for the KLOE and BaBar binning, we average the resulting smeared curve within each bin. We find that the required benchmark parameters do not depend strongly on the choice of data set. As we will discuss in Section \ref{sec:constraints}, the choice of such a suppressed lepton coupling relative to the hadronic coupling additionally arises from requiring consistency with the strong constraints on leptonic dark photon decays imposed by the BaBar experiment \cite{BaBar:2014zli} as well as the electron anomalous magnetic moment $g_e-2$. We show the experimental cross section data, $\Delta \sigma_{\mathrm{NP}}$, and inferred SM cross section for these benchmark points assuming the data is a representation of the total cross section in Fig.~\ref{fig:dsigma_low_gud}. To illustrate the shape of the cross section for a different choice of $Z'$ mass, we additionally show an alternative benchmark for KLOE with a mass of $m_{Z'}=0.82$ GeV in Fig.~\ref{fig:dsigma_KLOE_82}. For this plot, we employ couplings of $g_V^e=-2.7\times10^{-3}$ and $g_V^{ud}=0.1$.

Because there is an uncertainty on the discrepancy between the two HVP evaluations, we also present benchmark points in which we solve for a shift of $\Delta a_{\mu}^{\mathrm{HVP}}=0.76\times10^{-9}$. Again for $m_{Z'}=0.79$ GeV and $g_V^{ud}=0.1$, we find
\begin{align}\label{eq:unfixed_gud_1sig}
    g_V^e =
    \begin{cases}
        -2.0 \times 10^{-3} \;\;\; \mathrm{(CMD-2)} \\
        -1.7 \times 10^{-3} \;\;\; \mathrm{(KLOE)} \\
        -2.1 \times 10^{-3} \;\;\; \mathrm{(SND)} \\
        -1.9 \times 10^{-3} \;\;\; \mathrm{(BaBar)}
    \end{cases}
\end{align}

It should be noted that this $Z'$ mass
is close to the $\omega$ meson mass, $m_\omega \sim 782$~MeV, and that the $\rho$-$\omega$ mixing is relevant to explain the observed features in the hadronic cross section at these energies. In the next section, we discuss this consideration in more detail.

\subsection{Additional model considerations}

An examination of Figs.~\ref{fig:dsigma_low_gud} and \ref{fig:dsigma_KLOE_82} suggests that some parameter choices are more consistent with our understanding of SM photon physics than others: the shape of the $Z'$ peak means that there must be a complementary feature in the photon cross section, as there is no obvious feature present in the observed cross section. A peak in the photon spectrum below 1 GeV away from the $\rho$ and $\omega$ resonances has no presently-understood physical source. We examine two cases in which an unreasonable feature in the photon cross section may potentially be avoided. 

In the first case, illustrated in Fig.~\ref{fig:dsigma_low_gud}, the $Z'$ mass is 
close to the $\rho$ and $\omega$ resonance masses. The $Z'$ interference would affect the fits of the $\rho-\omega$ mixing in the vector meson dominance (VMD) framework~\footnote{ For an overview of VMD and $\rho-\omega$ mixing, see for example Ref.~\cite{OConnell:1995nse}.}, as an enhancement in the $\rho$ resonant feature such as the one shown in Fig.~\ref{fig:dsigma_low_gud} would require a greater $\omega$ interference to produce the sharp kink in the SM cross section. As the primary data employed in these fits of the mixing is the $\sigma(e^+e^- \to \mathrm{hadrons})$ data itself, one might attempt to adjust the strength of the $\rho$ resonance and the $\rho-\omega$ mixing. However,
the results of such analysis must also be compatible with the three pion production data, which directly constrains the $\omega$ and $\rho$ properties. Fits to these data prefer a smaller value for the $\rho$-$\omega$ mixing~\cite{Achasov:2003ir,BABAR:2021cde} than the two pion fits. As any modification from the $Z'$ in this energy region would require a larger $\rho-\omega$ mixing to reproduce the two-pion data, this scenario becomes a far-fetched possibility that is unlikely to be behind the explanation of the hadronic vacuum polarization discrepancy.

On the other hand, a $Z'$ mass further away from the $\rho$ and $\omega$ masses provides an unacceptable fit to the data.
Indeed, the shape shown in Fig.~\ref{fig:dsigma_KLOE_82} extends the $\omega$ feature beyond the $\omega$ mass region of 780 MeV and contains an additional sharp resonant feature at about 820 MeV, and thus presents an implausible scenario. 

Alternatively, one can reduce the $Z'$ effects by demanding it to reconcile the two hadronic vacuum polarization corrections at the one sigma level.   The result is that if a weaker correction to the cross section is demanded, the resonant feature become also understandably less prominent. However, the presence of the resonant feature and the requirement of the proximity of the
$Z'$ mass to the $\rho$ and $\omega$ masses remain in place, increasing the plausibility of this scenario only marginally.

\begin{figure}[h]
    \centering
    \includegraphics[width=0.9\textwidth]{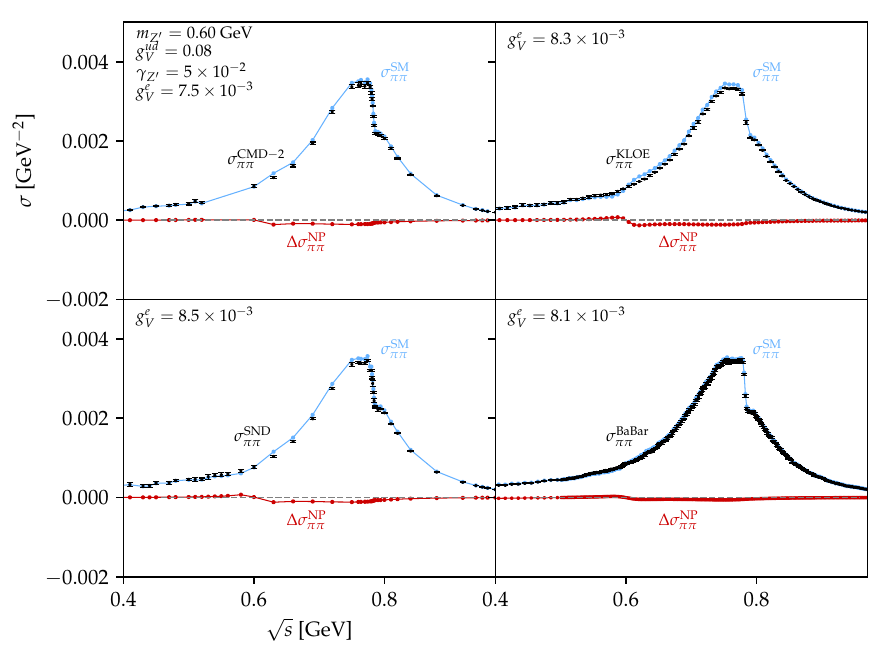} 
    \caption{The observed cross section for $e^+ e^- \to \pi^+\pi^-$ (black) from CMD-2 (top left), KLOE (top right), SND (lower left), and BaBar (lower right), compared to the change in cross section induced by a $Z'$ (red) with $m_{Z'}=0.60$ GeV, $g_V^{ud} = 0.08$, $\gamma=5\times 10^{-2}$, and $g_V^e=\{7.5, 8.3, 8.5, 8.1\}$ for CMD-2, KLOE, SND, and BaBar, respectively. For these benchmarks, we find $\Delta a_{\mu}^{\mathrm{HVP}} = 1.44\times10^{-9}$. We also show the resulting inferred SM cross section (blue).}
    \label{fig:dsigma_width}
\end{figure}

A more likely and alternative approach to the $Z'$-induced resonant feature is to increase the width of the $Z'$, thus suppressing the strength of the resonant peak. For the choice of parameters presented above, the value of $\gamma_{Z'}\equiv\Gamma_{Z'}/m_{Z'}$ is of order $(1-2)\times10^{-3}$ for $g_V^{ud}=0.1$; we find that a strong suppression of the $Z'$-induced feature may be obtained for $\gamma_{Z'}$ of the order of $5 \times 10^{-2}$, requiring a significant enhancement relative to the width arising only from pion and electron decays. The enhancement of the width is phenomenologically
challenging and demands, for instance, a new fermion or scalar charged under $Z'$ and with a mass below $m_{Z'}/2$. In addition, such a new particle  should not predominantly decay purely leptonically or to purely invisible final states in order to avoid experimental constraints such as the ones imposed by dark photon searches at the BaBar experiment. 

Here we present a possible realization of such an enhancement to the width. We shall assume that the $Z'$ gauge boson couples with a coupling of order one to a fermion $\psi_2$, and 
there is an additional fermion state $\psi_1$, which does not couple to $Z'$. The mass eigenstates are admixtures of these states, with the mixing being small, $\sin\theta_{\mathrm{mix}} \simeq \epsilon$ 
\begin{eqnarray}
\chi_2 & \simeq & \psi_2 + \epsilon \psi_1 \\
\chi_1 & \simeq & \psi_1 -\epsilon \psi_2,
\end{eqnarray}
where $m_{Z'} > 2 \ m_{\chi_2} > 2 \ ( m_{\chi_1} + m_\pi)$, and $m_\pi$ is the neutral pion mass. Under these circumstances, $Z'$ will
decay promptly to a pair of $\chi_2$ states.  Due to the mixing, which induces a small $Z'_\mu \bar{\chi}_2 \gamma_\mu \chi_1$ coupling, the $\chi_2$ state will subsequently  decay into pions and a $\chi_1$ state (it would decay into an electron positron pair, instead of pions, if the pion channels were kinematically closed). 

We require that the $Z'$
decay mode into $\chi_2$ pairs is the dominant one, so that the $Z'$ production
will lead to pions and missing energy in the final state, which fulfills the phenomenological requirement stated above. The width of 
the $Z'$ decay into $\chi_2$ will be given by
\begin{equation}
\Gamma(Z' \to \chi_2 \chi_2) = m_{Z'} N_2 \frac{g_2^2}{12 \pi} \sqrt{ 1 - \frac{4 m_{\chi_2}^2}{m_{Z'}^2} } \left(1+\frac{2 m_{\chi_2}^2}{m_{Z'}^2}\right),
\end{equation}
where $N_2$ denotes the $\chi_2$ (and $\chi_1$) multiplicity.        
For definiteness, let us assume that $\epsilon \sim 10^{-3}$, the $\chi_1$ mass
is of order 10~MeV, while the $\chi_2$ mass is about 250 MeV. Due to the smallness of $\epsilon$, the $Z'$ coupling to $\chi_1$ and therefore the invisible decay width of $Z'$ is sufficiently suppressed. In order to enhance the total decay width to obtain $\gamma = 5 \times 10^{-2}$, we must
demand
\begin{equation}
    N_2 \ g_2^2 \simeq 2.  
\end{equation}
Therefore, one can obtain a plausible scenario by choosing reasonable, 
perturbative couplings where $g_2^2/(4 \pi) \lesssim 0.16/N_2$.  

The suppression of the resonant feature in this case allows for more flexibility in the choice of $Z'$ mass and couplings and hence a more realistic scenario. In Table~\ref{tab:gamma_benchmarks}, we show a few example benchmarks with $\Delta a_{\mu}^{\mathrm{HVP}}=1.44\times10^{-9}$; note that because the increased width suppresses the strength of the peak near the $Z'$ mass, one requires larger values for $|g_V^e|$. We present masses both above and below the $\rho$ resonance; note that due to electroweak precision considerations, the lower masses of $m_{Z'}\lesssim 0.7$~GeV are preferred over those above the resonance \cite{Keshavarzi:2020bfy}. Masses lighter than around $0.6$~GeV start to come into tension with BaBar leptonic decay and pion mass bounds, which will be discussed in more detail in the next section. In Fig.~\ref{fig:dsigma_width}, we show the cross sections for a choice of $\gamma_{Z'} = 5\times 10^{-2}$, with $m_{Z'}=0.60$ and $g_V^{ud}=0.08$. 

\begin{figure}[h]
    \centering
    \includegraphics[width=0.9\textwidth]{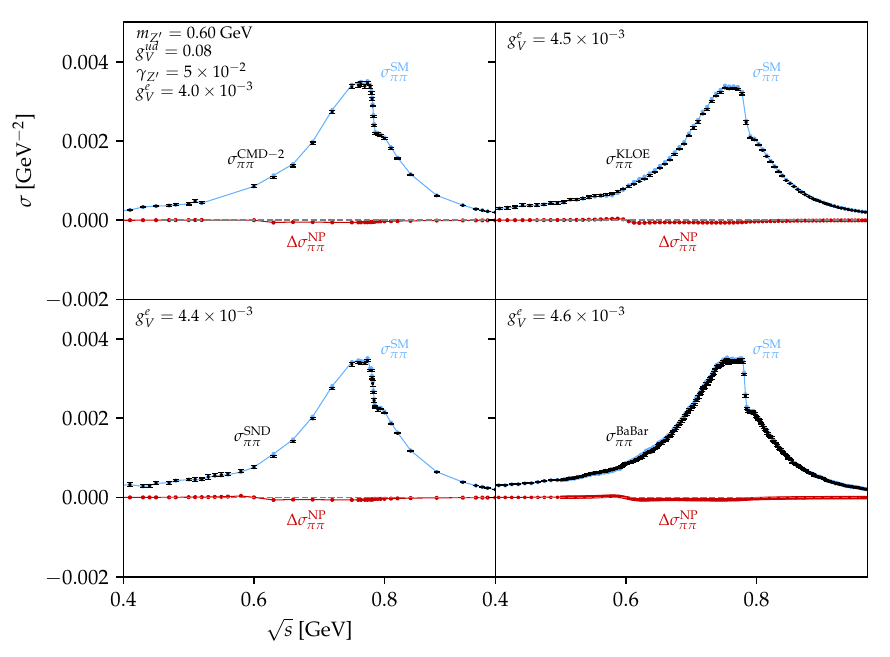} 
    \caption{The observed cross section for $e^+ e^- \to \pi^+\pi^-$ (black) from CMD-2 (top left), KLOE (top right), SND (lower left), and BaBar (lower right), compared to the change in cross section induced by a $Z'$ (red) with $m_{Z'}=0.60$ GeV, $g_V^{ud} = 0.08$, $\gamma=5\times 10^{-2}$, and $g_V^e=\{4.0, 4.5, 4.4, 4.6\}$ for CMD-2, KLOE, SND, and BaBar, respectively. For these benchmarks, we find $\Delta a_{\mu}^{\mathrm{HVP}} = 0.76\times10^{-9}$. We also show the resulting inferred SM cross section (blue).}
    \label{fig:enhanced_1sigma}
\end{figure}

As an illustration, we also calculate example benchmark values for resolving the tension at $1\sigma$, solving for $\Delta a_{\mu}^{\mathrm{HVP}} \simeq 0.76 \times10^{-9}$. For $m_{Z'}=0.6$~GeV and $g_V^{ud}=0.08,$
\begin{align}\label{eq:enhanced_1sigma_benchmarks}
    g_V^e =
    \begin{cases}
        4.0 \times 10^{-3} \;\;\; \mathrm{(CMD-2)} \\
        4.5 \times 10^{-3} \;\;\; \mathrm{(KLOE)} \\
        4.4 \times 10^{-3} \;\;\; \mathrm{(SND)} \\
        4.6 \times 10^{-3} \;\;\; \mathrm{(BaBar)}
    \end{cases}
\end{align}
We show the resulting curves for these benchmarks in Fig.~\ref{fig:enhanced_1sigma}.

Let us add in closing that although in this scenario there will be a $Z'$ resonant contribution to the two pion final states associated with its decays into $\chi_2$ states, such a contribution is suppressed by $(g_V^e)^2$ and becomes small with respect to the one induced by the mixing of the photon and $Z'$ with the QCD meson states, considered above. \\

\begin{table}
    \centering
    \begin{tabular}{|c||c|c|c|c|c|c|}
    \hline
    & \multirow{2}{*}{$m_{Z'}$ (GeV)} & \multirow{2}{*}{$g_V^{ud}$}  & \multicolumn{4}{c|}{$g_V^e \times 10^{3}$} \\
    Benchmark &  & & CMD-2 & KLOE & SND & BaBar \\ \hline
    1 & 0.60 & 0.08 & $7.5$ & $8.3$ & $8.5$ & $8.1$  \\ \hline
    2 & 0.60 & 0.07 & $8.6$ & $9.4$ & $9.7$ & $9.2$ \\ \hline
    3 & 0.65 & 0.08 & $7.6$ & $7.8$ & $7.1$ & $7.7$ \\ \hline
    4 & 0.65 & 0.07 & $8.6$ & $8.9$ & $8.1$ & $8.7$ \\ \hline
    5 & 0.65 & 0.06 & $10.0$ & $10.4$ & $9.4$ & $10.2$ \\ \hline
    6 & 0.90 & 0.09 & $-8.0$ & $-8.3$ & $-7.9$ & $-8.2$ \\ \hline
    7 & 0.90 & 0.08 & $-9.0$ & $-9.4$ & $-8.9$ & $-9.2$ \\ \hline
    \end{tabular}
    \caption{Example benchmark points with an enhanced width of $\gamma_{Z'}=5\times10^{-2}$ that satisfy $\Delta a_{\mu}^{\mathrm{HVP}} = 1.44 \times 10^{-9}$ for the respective listed experiment.}
    \label{tab:gamma_benchmarks}
\end{table}

\section{Constraints}
\label{sec:constraints}

There are a number of experimental bounds that constrain this scenario, arising from flavor physics, collider physics, and precision measurements. We provide a summary and discussion of these constraints below. \\

\textbf{Electron $g-2$:} precision measurements of the electron anomalous magnetic moment~\cite{Odom:2006zz}, which will receive contributions from $Z'$ loops. This places a mass-dependent bound on the value of $|g_V^e|$, and becomes more constraining for lighter $Z'$ masses. We find that~\cite{Pospelov:2008zw} 
\begin{align}
\Delta a_{e}^{Z',\mathrm{loop}} \sim (4.0 \times 10^{-13}) \left( \frac{600~\mathrm{MeV}}{m_{Z'}}\right)^2 \left( \frac{g_V^e}{8 \times 10^{-3}} \right)^2
\end{align}

The current bound on $a_e$ depends on the comparison of the theoretical predictions~\cite{Aoyama:2019ryr,Volkov:2019phy} and the most recent experimental determination~\cite{Fan:2022eto}, but is also affected by the current uncertainty in the value of the electromagnetic structure constant~\cite{Parker:2018vye,Morel:2020dww}. An approximate bound may be set,
\begin{equation}
 |\Delta a_{e}^{Z',\mathrm{loop}} | \lesssim \  10^{-12}.
\end{equation}
For masses of $0.6$ GeV, we find that this translates to a bound of $|g_V^e| \lesssim 12.5 \times 10^{-3}$, which is satisfied for the relevant benchmark couplings shown in Table~\ref{tab:gamma_benchmarks}. A resolution of the outstanding theoretical questions about the electron $g-2$ and an order of magnitude improvement of the experimental precision would allow one to probe the lighter-mass benchmarks presented in Table \ref{tab:gamma_benchmarks}.
Meanwhile, for a $Z'$ mass of order 0.9~GeV, the bound on $|g_V^e|$ is on the order of $19\times10^{-3}$.\\

\noindent \textbf{BaBar leptonic decay}: seach for $e^+e^- \to Z' \gamma$, $Z' \to e^+ e^-$ ~\cite{BaBar:2014zli}. This places a bound on the $Z'$ coupling $|g_V^e|$; because this bound depends on the respective branching fractions of the $Z'$ decays, the precise bound depends on the model interpretation. 

The BaBar bound on the mixing of a dark photon $A'$ with the SM photon is approximately $\epsilon_{A'} \lesssim 10^{-3}$ for $m_{A'} \simeq 0.8$ GeV and $\epsilon_{Z'} \lesssim 6\times 10^{-4}$ for $m_{A'} \simeq 0.6$ GeV. The bound on the mixing $\epsilon$ translates to an upper bound of approximately $g_{\mathrm{lim}}^e \simeq 3.3 \times 10^{-4}$ and $2.0 \times 10^{-4}$ for $m_{Z'}=0.8$ and 0.6 GeV, respectively. In the dark photon analysis presented in Ref.~\cite{BaBar:2014zli}, the branching ratios are calculated assuming that the $A'$ picks up charge-proportional couplings to the quarks, electron, and muon through kinetic mixing. 

In the case examined here, the large hierarchy between the quark and lepton couplings, as well as the enhancement of the width through additional decays, means that the branching ratio to electrons is heavily suppressed. As such, one must rescale the bound on $(g_V^e)^2$ by the ratio of the branching ratio to electrons in the two respective models, evaluating the inequality $(g_V^e)^2 \lesssim (g_{\mathrm{lim}}^e)^2 \mathrm{BR}_{A' \to ee} / \mathrm{BR}_{Z' \to ee}(g_V^e, g_V^{ud}, \gamma_{Z'})$.
One finds a resulting bound for the $Z'$ model of $|g_V^e| \lesssim 5.5 \times 10^{-3}$ for $g_V^{ud}=0.1$ with unenhanced width and $m_{Z'}\simeq 0.8$ GeV, and $|g_V^{e}| \lesssim 1.2\times 10^{-2}$ for $\gamma_{Z'}=5\times10^{-2}$ and $m_{Z'}\simeq0.6$ GeV. An examination of the example parameters shown in Table~\ref{tab:gamma_benchmarks} thus indicates that one cannot take values of $g_V^{ud}$ lower than about 0.06, which would be preferred by pion mass constraints, without violating BaBar bounds due to the larger required values of $g_V^e$. Note that the dark photon analysis presented by BaBar assumes a narrower width than one obtains in this $Z'$ model; accounting for this difference would further weaken the bounds, and we therefore do not undertake a more detailed analysis in this direction.\\

\noindent \textbf{BaBar invisible decay}: search for $e^+e^- \to Z' \gamma$, $Z' \to$ invisible \cite{BaBar:2017tiz}. This process places a bound on $g_V^e$ as long as the new boson has relevant purely invisible decays. In our proposed model, the $Z'$ either does not include any couplings to invisible particles or decays to final states that include pions alongside missing energy, which does not pass the selection criteria for this search. \\

\noindent \textbf{Belle-II}: search for $e^+ e^- \to \mu^+ \mu^- Z'$ with invisible $Z'$ decays \cite{Belle-II:2019qfb}. This is a similar bound to that from  BaBar, but also includes a bound on the coupling to the muon, depending on whether $Z'$ decays to an invisible final state. In our model, we do not include a coupling to muons.\\

\noindent \textbf{LEP2}: measurement of $e^+ e^- \rightarrow q\bar{q}$ \cite{ALEPH:2013dgf}. This places an effective bound on $\tilde{g}$, as the process is sensitive to $g_V^e$ and $g_V^q$. This bound can be avoided for small enough $|g_V^e g_V^q|$~\cite{DiLuzio:2021uty}. 
Since the measurement does not distinguish between light quarks, one may express this bound in a more precise way, by including all gauge boson contributions and demanding that the variation of the cross section is smaller than about one percent, which is the characteristic precision of the LEP2 hadronic cross section measurement. Although all quarks, apart from the top quark, can be produced in pairs at LEP2, we shall conservatively concentrate on the production of up and down quarks, which are the ones that couple in a relevant way to the new gauge boson.

\begin{figure}
    \centering
    \includegraphics[width=0.6\textwidth]{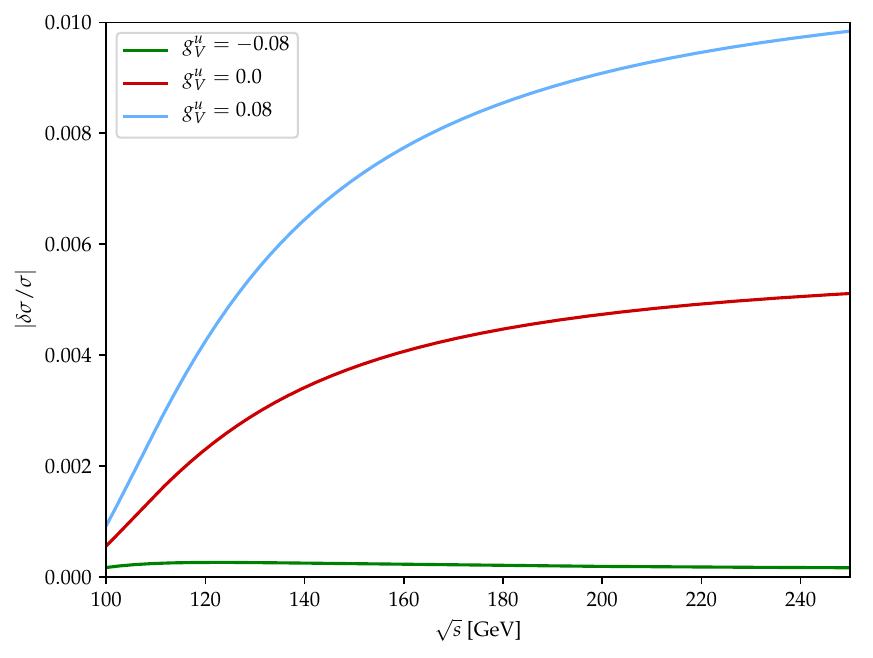}
    \caption{Relative variation of the hadron cross section at LEP2 due to the contribution of the new gauge boson, as a function of the  center of mass energy, for $g_V^{ud}=0.08$, $g_V^e=9.0\times10^{-3}$, and values of $g_V^u=-0.08$ (green), 0.0 (red), 0.08 (blue).}
    \label{fig:deltasigLEP2}
\end{figure}

The LEP2 center of mass energies are high enough that the gauge boson mass $m_{Z'}$, as well as the gauge boson width effects, may be neglected in this analysis.
Defining the chiral coupling
\begin{equation}
D_f =  \frac{g}{\cos\theta_W}(T_3^f - Q_f \sin^2\theta_W),
\end{equation}
and 
\begin{equation}
    f_Z = \frac{s}{s-M_Z^2},
\end{equation}
where $T_3^f$ and $Q_f$ are the chiral fermions weak isospin and electromagnetic charges,
the different hadronic cross section contributions are proportional to
\begin{eqnarray}
\sigma_Z & \propto & 0.25 f_Z^2 (D_{e_L}^2+D_{e_R}^2) \sum_{q=u,d}(D_{q_L}^2+D_{q_R}^2),
\nonumber\\
\sigma_\gamma & \propto & e^4 Q_e^2 ( Q_u^2 +Q_d^2),
\nonumber\\
\sigma_{Z'} & \propto & (g_{V}^e)^2 \left((g_{V}^u)^2+(g_V^d)^2\right),
\nonumber\\
\sigma_{Z\gamma} & \propto & 0.5 f_Z e^2 Q_e (D_{e_L} + D_{e_R}) \sum_{q=u,d} Q_q (D_{q_L}+D_{q_R}),
\nonumber\\
\sigma_{Z' Z} & \propto &  0.5 f_Z g_V^e  (D_{e_L} + D_{e_R}) \sum_{q=u,d} g_V^q (D_{q_L}+D_{q_R}),
\nonumber\\
\sigma_{Z'\gamma} & \propto & 2 e^2 Q_e g_V^e  \sum_{q=u,d} Q_q g_V^q.
\end{eqnarray}
In the above, $\sigma_Z$, $\sigma_\gamma$,$\sigma_{Z'}$,$\sigma_{Z\gamma}$,$\sigma_{Z'Z}$ and $\sigma_{Z'\gamma}$ represents the $Z$, $\gamma$ and $Z'$ contributions as well as the $Z\gamma$, $Z'Z$ and $Z'\gamma$ interference contributions, respectively.
The phenomenological requirement may be expressed as
\begin{equation}
    \frac{\delta\sigma}{\sigma} = \frac{\sigma_{Z'\gamma}+\sigma_{Z'Z}+\sigma_{Z'}}{\sigma_Z+\sigma_{Z\gamma}+\sigma_\gamma} \lesssim 0.01.
    \label{eq:LEP2bound}
\end{equation}
In Fig.~\ref{fig:deltasigLEP2} we plot the absolute value of the relative variation of the cross section for benchmark scenario $g_V^{ud}=0.08$, $g_V^e=9.0\times10^{-3}$ for values of $g_V^u = -g_V^{ud},0,$ and $g_V^{ud}$.  As can be seen from this figure, the variation becomes larger for larger values of
$g_V^u$, but stays at values lower than the LEP2 bound, Eq.~(\ref{eq:LEP2bound}), for $g_V^u = -g_V^{ud}$.

In order to understand qualitatively the behavior of $\delta \sigma/\sigma$, one can neglect the $Z$ contribution. In this case, and ignoring the small $\sigma_{Z'}$ contribution, the LEP2 bound can be rewritten as
\begin{equation}
2 |g_V^e| \frac{| Q_u g_V^u + Q_d g_V^d|}{e^2 (Q_u^2 + 
 Q_d^2)} \lesssim 0.01,
\end{equation}
or, equivalently
\begin{equation}
 \left(\frac{|g_V^e|}{2 \times 10^{-3}}\right) \left|g_V^u + g_V^{ud} \right| \lesssim 0.5.
 \label{eq:LEP2appr}
\end{equation}
As can be seen from Eq.~(\ref{eq:LEP2appr}) and also shown for the complete expression in Fig.~\ref{fig:deltasigLEP2}, one can satisfy
this bound by relating the up and down couplings such that the modifications to the individual up- and down-quark production rates cancel one another.  The $Z$ contribution introduces an energy dependence of $\Delta \sigma/\sigma$ that generically weakens the LEP2 bound for values of $g_V^u$ away from the cancellation region $g_V^{ud} \approx - g_V^u$.

Observe that  the couplings required to resolve the discrepancy between the BMW lattice-QCD and data-driven calculations can be close to the limit imposed by LEP2, depending on $g_V^u$, making this channel a potential method for probing such a $Z'$. Future $e^+e^-$ colliders such as FCC-ee, which would provide improved precision on the $e^+ e^- \to q\bar{q}$ rate, present an opportunity to search for this new physics.\\

\noindent \textbf{Neutrino-electron scattering}: observations of $\nu_e e^- \to \nu_e e^-$, and other neutral current variations on this process, from Borexino \cite{Bellini:2011rx}, TEXONO \cite{PhysRevD.81.072001,TEXONO:2006xds}, and CHARM II \cite{CHARM-II:1994dzw}, which restricts the product $g_V^e g_V^{\nu} \lesssim 10^{-6}$ (see e.g. Ref.~\cite{Bilmis:2015lja}). In our model, we do not induce a neutrino coupling, so these bounds are avoided. \\

\noindent \textbf{Flavor changing meson decays:} measurements of the flavor-changing decays of mesons, including $B \to K \bar{\nu} \nu$, $B \to K e^+ e^-$ , $K^+ \to \pi^+ \bar{\nu} \nu$, and $B \to K \pi^+ \pi^-$~\cite{BaBar:2013npw,E949:2008btt,HFLAV:2008dnl}. Since the quark couplings of the $Z'$ are of order 0.1, meson decay bounds may be relevant. In particular, one may have penguin diagrams where the initial bottom/strange quark is converted to an up quark through a W loop, with the up then radiating a $Z'$. However, three considerations reduce the strength of these bounds for our model. Firstly, the processes $B \to K \bar{\nu} \nu$ and $K^{+} \to \pi^{+} \bar{\nu} \nu$ can probe new physics if the new physics decays invisibly or is long-lived; with the choice of couplings included in this model, the $Z'$ decays visibly and is not long-lived. Secondly, the contribution to the process $B\to K e^+e^-$ is suppressed by the small branching ratio of the $Z'$ to leptons. 

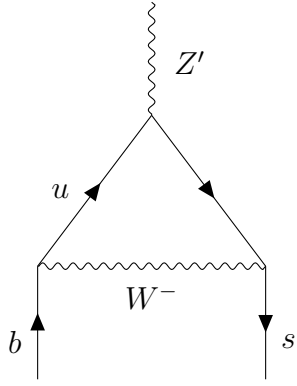
\begin{figure}[t] 
\centering
\begin{tikzpicture}
\begin{feynman}
\vertex (f1);
\vertex[above=1.5cm of f1] (a);
\vertex[right=1.5cm of a] (x1);
\vertex[above=2cm of x1] (f2);
\vertex[above=1.5cm of f2] (b);
\vertex[right=3cm of a] (c);
\vertex[right=3cm of f1] (f3);
\diagram*{
(f1)--[fermion] (a) --[fermion] (f2) --[fermion] (c) --[fermion] (f3);
(f2)--[boson] (b);
(a)--[boson] (c);
};
\node at (-0.3, 0.5) {$b$};
\node at (3.3, 0.5) {$s$};
\node at (0.3, 2.5) {$u$};
\node at (2., 4.2) {$Z'$};
\node at (1.5, 1.1) {$W^-$};
\end{feynman}
\end{tikzpicture}
\caption{Example penguin diagram for $B\to K$ flavor-changing processes involving a radiated $Z'$.}
\end{figure}

This leaves the $B \to K \pi^+ \pi^-$ processes as the final consideration. To evaluate the contribution arising from $Z'$ diagrams, we rescale the $B \to K \gamma$ branching ratio by the appropriate factors relative to the SM case, as this process is the analogous SM process to the $Z'$ penguin diagram, and compare with the measured branching fraction $BR(B \to K \pi^+ \pi^-) \simeq 5 \times 10^{-5}$.
Because the $Z'$ does not couple to the heavier quarks in this model, the rate of $Z'$ processes is suppressed by a factor of $|V_{ub}|^2 |V_{us}|^2 = 6.7\times 10^{-7}$ relative to the SM processes with $|V_{tb}|^2 |V_{ts}|^2 = 1.7\times 10^{-3}$. Additionally, these processes are associated to dipole operators that are therefore chirally suppressed by powers of the light quark masses to the charged gauge boson mass, thus picking up an additional factor of $(m_u/m_W)^2 \simeq 10^{-9}$. Rescaling $BR(B \to K \gamma)\simeq 3.5 \times 10^{-4}$ by these factors and comparing with the measured branching ratio of $BR(B \to K \pi^+ \pi^-) \simeq 5 \times 10^{-5},$ we find that flavor violating constraints are insignificant for the present scenario. \\

\noindent \textbf{Isospin breaking processes:} measurements and lattice-QCD calculations of the pion mass splitting, $\Delta m_{\pi} \equiv m_{\pi^+} - m_{\pi^0}$, place bounds on the allowed values of $|g_V^u - g_V^d|$. However, due to the fact that the lattice-QCD simulations are performed for the case of the massless photon~\cite{Gagliardi:2021vpv} and the $Z'$ mass in our case is of order of 600~MeV, the bounds should be reevaluated considering the gauge boson mass suppression effects. Rescaling the mass splitting computed in the lattice by a simple mass factor $(m_\pi/m_{Z'})^2$ leads to 
values of $\Delta m_\pi \simeq 0.018 \times (g_V^{ud}/0.1)^2$~MeV,
which for $g_V^{ud} \simeq 0.1$ is within the order of the uncertainty in the lattice-QCD
computation of this quantity.

Alternatively, one can estimate the effect of the isospin breaking interactions by taking the mean  electromagnetic interactions effect and replacing the Coulomb potential $e^2/r$ by a Yukawa one $(g_V^{ud})^2 e^{-mr}/r$. Comparing the relevant factors in the two potentials, one finds that the $Z'$ shift can be approximated by $(4.6~\mathrm{MeV})\times (g_V^{ud})^2 e^{-mr}/e^2$. Taking the pion charge radius as the characteristic quark separation and using $m \simeq 0.6$~GeV, this leads to a pion mass shift of order $(0.044-0.056) \times (g_V^{ud}/0.1)^2$~MeV, depending on whether one uses $r_{\pi} = 0.66$~fm \cite{Cui:2021aee} or $r_{\pi} = 0.74$~fm \cite{A1:1999kwj}. For $g_V^{ud} \simeq 0.1$, these results imply a shift that is within the $\Delta m_\pi$ lattice-QCD computation uncertainty~\cite{Gagliardi:2021vpv}.

Finally, one may instead employ the Cottingham method \cite{Cottingham:1963zz,Donoghue:1996zn,Stamen:2022uqh,Crivellin:2022gfu} to estimate the effect of the $Z'$ on the pion mass splitting. In this approach, which is more reliable than the previous estimates, we calculate the contribution from the $Z'$ as~\cite{Crivellin:2022gfu}
\begin{align}
    \Delta m_{\pi}^2\Big|_{Z'} = \frac{(g_V^{ud})^2}{32\pi^2} \int_0^{\infty} ds \frac{s}{s+m_{Z'}^2} (F_{\pi}^V(-s))^2 \left( 4W + \frac{s}{m_{\pi}^2}(W-1) \right),
\end{align}
where $W=\sqrt{1+4m_{\pi}^2/s}$. We employ an approximate analytical expression for $F_{\pi}^V(-s) =(A_{\rho \gamma}/e) g_{\rho \pi}/(s+m_{\rho}^2)$ to evaluate the integral, where $A_{\rho \gamma}$, which will be discussed in more detail in the $\pi^+\pi^-$ scattering analysis, parametrizes the mixing of the photon with the $\rho$. While this expression for $F_{\pi}^V$ is not normalized to $F_{\pi}^V(0)=1$, it provides a decent approximation of the integral as the infrared  provides a subdominant contribution. We find a bound of $g_V^{ud} \lesssim 0.09$ for $m_{Z'}=0.9$ GeV and $g_V^{ud}\lesssim0.08$ for $m_{Z'} = 0.6$ GeV, in approximate agreement with Ref~\cite{Crivellin:2022gfu}.

Note that these bounds become more stringent for lighter $Z'$ masses. Although these bounds may be avoided by decreasing $g_V^{ud}$, this consequently requires an increase in $g_V^e$, which is bounded by the electron $g-2$ and which also becomes more constrained for lighter $Z'$ masses. These bounds therefore create preference for heavier $Z'$ masses. The range of masses around $0.6-0.7$ GeV are able to satisfy these bounds while remaining consistent with the electroweak precision measurements, which prefer $m_{Z'} \lesssim 0.7$ GeV.

While these calculations provide an estimate of the effect of the $Z'$ isospin breaking interactions on $\Delta m_{\pi}$ and the order of magnitude of the bounds on $g_V^{ud}$, a lattice-QCD calculation is in order to provide a more precise evaluation. \\

\noindent \textbf{$\pi^+ \pi^-$ scattering:} measurements of the cross section for $\pi^+ \pi^- \to \pi^+ \pi^-$ scattering place bounds on the allowed values of $g_V^{ud}$. For the mass region we are interested in, we compare with data from Refs.~\cite{Baillon:1972tx,Protopopescu:1973sh}; within the center-of-mass energy range close to the $Z'$ mass, we expect the contribution from the $Z'$ to induce a feature similar to that induced in the $e^+e^- \to \pi^+ \pi^-$ scattering. In this case, the contribution is not suppressed by a factor of $g_V^e$ and is instead proportional to $(g_V^{ud})^2$, making this a particularly relevant bound. 

\begin{figure}[h!]
    \centering
    \includegraphics[width=0.3\textwidth]{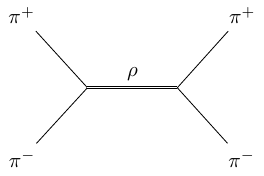}
    \includegraphics[width=0.45\textwidth]{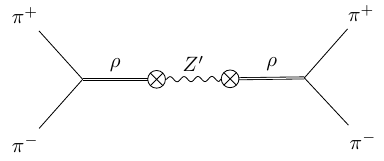}
    \caption{Diagrams for $\pi^+\pi^- \to \pi^+\pi^-$ scattering with a $\rho$ mediator (left) and $\pi^+ \pi^-$ scattering with a single $Z'$ insertion (right).}
    \label{fig:my_label}
\end{figure}

To gain a rough sense of the impact a $Z'$ feature would have on the $\pi^+\pi^-$ scattering cross section, we employ the following formalism, which is similar to that discussed in Ref~\cite{Fang:2021wes}. The propagator for the $\rho$,
\begin{align}
    D(\rho) = \frac{1}{s - m_{\rho}^2 + i m_{\rho}\Gamma_{\rho}},
\end{align}
will be modified by insertions of the $Z'$, leading to a series in powers of $Z'$ insertions
\begin{align}
    D(\rho) &\to D(\rho) + D(\rho) A_{\rho Z'} D(Z') A_{\rho Z'} D(\rho) + ... \\
    &= \sum_{n=0}^{\infty} D(\rho) (A_{\rho Z'}^2 D(Z') D(\rho))^n \\
    &= \frac{D(\rho)}{1-A_{\rho Z'}^2 D(Z') D(\rho)},
\end{align}
where $A_{\rho Z'}$ is the mixing between the $Z'$ and the $\rho$. We also account for $\rho-\omega$ mixing, which includes similar diagrams as the $Z'-\rho$ mixing, but which also leads to the addition of diagrams with insertions of both $Z'$ and $\omega$. The above modification therefore becomes
\begin{align} \label{eq:Dpipi}
    D(\rho) &\to \sum_{n=0}^{\infty} D(\rho) (A_{\rho Z'}^2 D(Z') D(\rho) + A_{\rho\omega}^2 D(\omega) D(\rho))^n \\
    &= \frac{D(\rho)}{1-(A_{\rho Z'}^2 D(Z') D(\rho) + A_{\rho\omega}^2 D(\omega) D(\rho))}.
\end{align}

Taking the same approach as above for the $e^+ e^- \to \pi^+ \pi^-$ process, we find that the contribution involving the photon for this process may be written in terms of
\begin{align} \label{eq:Dphoton}
    & \; D(\gamma) A_{\rho \gamma} D(\rho) + D(\gamma) A_{\rho \gamma} D(\rho) A_{\rho \gamma} D(\gamma) A_{\rho \gamma} D(\rho) + ... \\
    =& \; \frac{D(\gamma) A_{\rho\gamma}  D(\rho)}{1-A_{\rho \gamma}^2 D(\gamma)D(\rho)}.
\end{align}
where $A_{\rho \gamma}$ is the mixing of the photon with $\rho$, and is related to $A_{\rho Z'}$ as $A_{\rho\gamma}=(\tfrac{e}{g_V^{ud}})A_{\rho Z'}$. 
As a check, we may first employ this formalism to reproduce the expression written in Eq.~(\ref{eq:ee_sig_ratio}). With the addition of a $Z'$, the higher-power insertions may include either a $Z'$ or a photon, leading to a common term in the denominator for the modified photon and $Z'$ propagators. This leads to a ratio between the two contributions of
\begin{align}
    \frac{g_V^e D(Z') A_{\rho Z'}}{- e D(\gamma) A_{\rho \gamma}}
\end{align}
where $g_V^e$ and $-e$ account for the respective vertices with the initial $e^+ e^-$. Thus, we find
\begin{align}
    \frac{\sigma_{\pi\pi}^{\mathrm{SM+Z'}}}{\sigma_{\pi\pi}^{\mathrm{SM}}} &= \left| 1 - \frac{g_V^e g_V^{ud}}{e^2} \frac{D(Z')}{D(\gamma)} \right|^2 \\
    &= \left| 1 - \frac{g_V^e g_V^{ud}}{e^2} \frac{s}{s-m_{Z'}^2 + im_{Z'}\Gamma_{Z'}} \right|^2
\end{align}
which agrees with the expression quoted in Eq.~(\ref{eq:ee_sig_ratio}). In taking this approach, we have made the simplifying assumption that the the mixing of the photon and $Z'$ with the $\omega$ also scale with $e$ and $g_V^{ud}$, respectively. 

We extract the mixing $A_{\rho \gamma}$ and $A_{\rho\omega}$ from the quantity $R(s)$, and find $A_{\rho\gamma}\simeq 0.033$ GeV$^2$ and $A_{\rho\omega} \simeq -0.006$ GeV$^2$; the details are contained in the Appendix. In fitting the observed $R(s)$ to extract mixing parameters, we have assumed SM physics; we have also checked the case where we account for the impact of the $Z'$ on $R(s)$, and find that the results of the $\pi^+\pi^-$ scattering analysis do not depend strongly on this assumption. We use the relation $A_{\rho Z'} = (\tfrac{g_V^{ud}}{e}) A_{\rho\gamma}$ to obtain $A_{\rho Z'}$ from $A_{\rho\gamma}$. 

\begin{figure}[h!]
    \centering
    \includegraphics[width=0.65\textwidth]{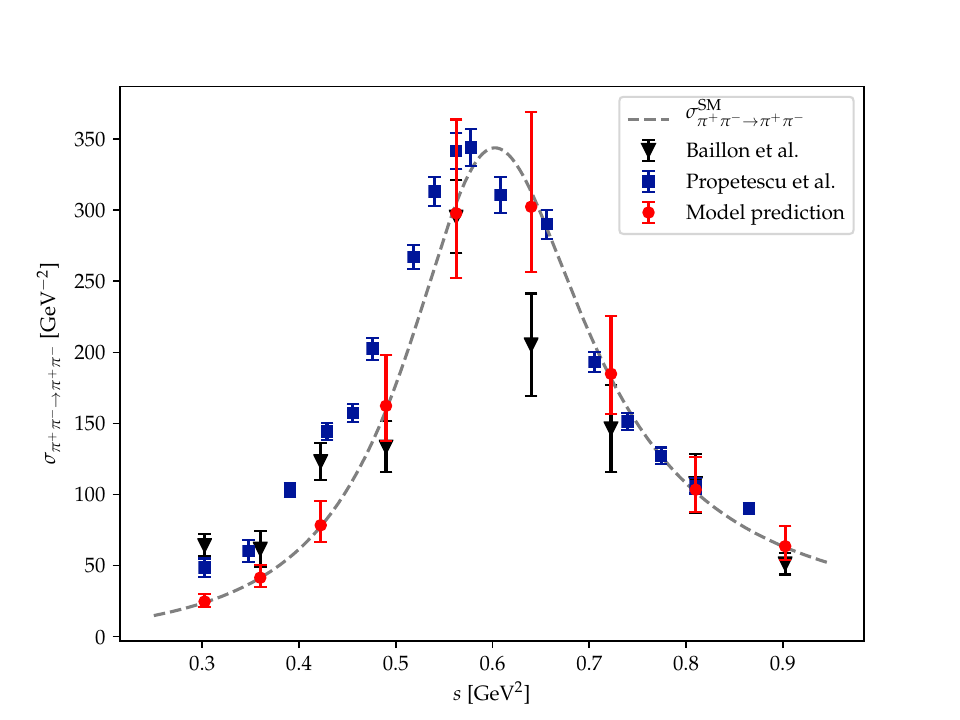}
    \caption{Comparison of data for $\pi^+\pi^- \to \pi^+\pi^-$ scattering with the expected cross section including $Z'$ modifications for $m_{Z'}=0.6$ GeV and $g_V^{ud}=0.1$. In this plot, the $Z'$ width is $\gamma_{Z'}=5\times10^{-2}$.}
    \label{fig:pipi}
\end{figure}

We may now solve for the modification of the $\rho$-mediated $\pi^+\pi^-$ scattering due to the $Z'$. The ratio of the modified $\rho$ propagator relative to the SM one is
\begin{align}
    \frac{D(\rho)^{\mathrm{SM+NP}}}{D(\rho)}(s) = \frac{1}{1-A_{\rho Z'}^2 D(Z') D(\rho) - A_{\rho \omega}^2 D(\omega) D(\rho)}
\end{align}
and thus the expected ratio of cross sections will be proportional to
\begin{align}
    \Delta_{\pi\pi}(s) \equiv \frac{\sigma_{\pi^+\pi^- \to \pi^+\pi^-}^{\mathrm{SM+NP}}}{\sigma_{\pi^+\pi^- \to \pi^+\pi^-}^{\mathrm{SM}}}(s) = \left| \frac{1}{1-A_{\rho Z'}^2 D(Z') D(\rho) - A_{\rho \omega}^2 D(\omega) D(\rho)} \right|^2.
\end{align}
As an illustration, we calculate $\sigma_{\pi^+\pi^- \to \pi^+\pi^-}^{\mathrm{SM}}$ as a vector-mediated scalar scattering process, focusing on the $s$-channel diagrams which will be most relevant near the $\rho$ resonance, finding the expression
\begin{align}
    \sigma_{\pi^+\pi^- \to \pi^+\pi^-}^{\mathrm{SM}}(s) = \frac{g_{\rho\pi}^4}{48\pi} \frac{s}{(s-m_{\rho}^2)^2 + \Gamma_{\rho}^2 m_{\rho}^2}.
\end{align}
In Fig~\ref{fig:pipi} we compare the expected cross section $\sigma_{\pi^+\pi^- \to \pi^+\pi^-}^{\mathrm{SM+NP}}$ in our model with the observed data; we account for a binning of 50 MeV in $\sqrt{s}$. We employ a value of $\alpha_{\rho\pi} = g_{\rho\pi}^2/4\pi = 2.7$ for the central values in Fig.~\ref{fig:pipi}, while the errorbars for the model prediction indicate the range of values for $\alpha_{\rho\pi} \in [2.4,2.9]$ \cite{Sakurai:1966zza}. We find that the deviations of our prediction from the data are on the order of the uncertainty in the data in the region of the $Z'$ and $\rho$ features. The authors of Ref.~\cite{Protopopescu:1973sh} emphasize that while they expect the qualitative behavior of their results to be correct, the quantitative values may have systematic deviations due to the extrapolation procedure; we conclude that the $\pi^+\pi^-$ scattering process does not rule out the presence of a $Z'$.

\section{Loop contributions to $a_\mu$} \label{sec:muong2}

The primary aim of this model is to reconcile experimental determinations of $\sigma_{\mathrm{had}}$ with the lattice-QCD calculation by the BMW collaboration. However, there is still a 1.6~$\sigma$ discrepancy between the value of $a_\mu$ obtained from the BMW lattice-QCD calculation of the hadronic vacuum polarization effects and the measured value for $a_\mu$. 
Moreover, the large hierarchy between the values of $g_V^e$ and $g_V^{ud}$ required to obtain the necessary shift  $\Delta a_{\mu}^{\mathrm{HVP}}$ while also satisfying experimental constraints raises the question of why these couplings differ by orders of magnitude. One possibility is that only the quark couplings appear in the Lagrangian at tree level,  with the couplings
to other SM particles  induced by a kinetic mixing $\epsilon$ between the $Z'$ and the photon. This mixing could be
partially induced by a loop contribution of the up and down quarks, which are charged under $Z'$ and electromagnetic interactions.
Thus, interactions between the new $Z'$ and the electron are induced at the loop level and are suppressed relative to the quark couplings. Note that in this case, a small coupling to the muon is also induced:
\begin{equation}
g_V^e = g_V^\mu.
\end{equation}

It is then reasonable to ask whether this model under the above assumptions may reconcile BMW's lattice-QCD calculation with the experimental measurement.
In particular, the contribution from $Z'$ loops gives a modification of~\cite{Pospelov:2008zw}
\begin{align}
\Delta a_{\mu}^{Z',\mathrm{loop}} = (1.56 \times 10^{-8}) \left(\frac{600 \; \mathrm{MeV}}{m_{Z'}} \right)^2 \left( \frac{g_V^e}{8 \times 10^{-3}} \right)^2.
\end{align}
We can compare this to the discrepancy between the lattice-QCD BMW determination and the experimentally-measured value of $a_{\mu}$, which is approximately $\Delta a_{\mu}^{\mathrm{BMW-exp}} \simeq 1\times 10^{-9}$. For the benchmark values presented in Eqs.~(\ref{eq:unfixed_gud_benchmarks}), the predicted correction would be of order $0.9\times10^{-9}$ to $1.5\times10^{-9}$. However, for the more realistic enhanced-width cases presented in Table~\ref{tab:gamma_benchmarks}, the correction is of order $10^{-8}$; for the relaxed $1\sigma$ benchmark in Eq.~\ref{eq:enhanced_1sigma_benchmarks}, meanwhile, the correction is of order $5\times10^{-9}$. Hence, the explanation of the small $g_V^e$ coupling proceeding from a kinetic mixing with the photon becomes unacceptable in this case.

\section{Conclusions}
\label{sec:conclusions}
In this article we study the possibility of including new physics that affects the Standard Model hadronic cross section $\sigma(e^+e^-\to { \mathrm{hadrons}})$ in order to reconcile the hadronic vacuum polarization contributions computed from dispersion relations and the one recently obtained by the lattice-QCD BMW collaboration. This may be done by introducing a new vector that interferes destructively with the photon-induced cross section. We showed that a successful model demands a large hierarchy between the quark and lepton couplings. Most phenomenological constraints may be fulfilled in such a case, although the presence of resonant features in the hadronic production spectrum demands the $Z'$ to be close in mass to the $\rho$
and $\omega$, making the scenario implausible. A more realistic scenario is obtained when the resonant feature is avoided due to  an enhanced width, and we described a possible realization of such scenario. For the parameter values considered in this work, improved precision from a future $e^+e^-$ collider and improved determinations of $a_e$ would allow one to probe this scenario. While we have employed pre-existing results for the tension in the $a_{\mu}$ theory prediction and for the experimental data, more precise measurements will concretely determine the required benchmark values.

The model we described may be understood as a low energy effective theory. Reconciling this effective theory with the electroweak interactions is challenging due to the required isospin breaking couplings of the $Z'$ to up and down quarks,  which are not consistent with the $SU(2)_L\times U(1)_Y$ gauge symmetry of the SM and hence should appear as  effective low energy interactions after electroweak symmetry breaking\footnote{Note that in the absence of an ultraviolet completion, one obtains infrared enhanced contributions to the decays of the $W$ and $Z$ bosons, which provide a bound on $g_V^{q}$ of the same order as that from the pion mass splitting; see e.g. \cite{Karshenboim:2014tka}}. 

~\\
~\\
\textbf{Acknowledgements} \\

C.W.\ would like to thank the Aspen Center for Physics, which is supported by National Science Foundation grant No.~PHY-1607611, where part of this work has been done. We would like to thank R. Boughezal, G. Bodwin,  T. Hobbs, F. Petriello and, in particular, F. Herren and R. van de Water for useful discussions and comments. We would also like to thank M. Hoferichter, A. Crivellin, and L. Darm\'{e} for interesting comments. C.W.\ has been partially supported by the U.S.~Department of Energy under contracts No.\ DEAC02- 06CH11357 at Argonne National Laboratory. The work of C.W.\ and N.C.\ at the University of Chicago has also been supported by the DOE grant DE-SC0013642.

\newpage
\appendix
\section{$\rho-\omega$ mixing}

\begin{figure}[h!]
    \centering
    \includegraphics[width=0.35\textwidth]{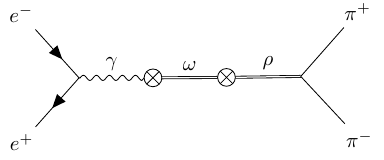}
    \includegraphics[width=0.5\textwidth]{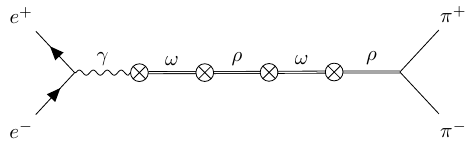}
    \caption{Leading diagrams for $e^+ e^- \to \pi^+ \pi^-$ that include $\omega$ insertions.}
    \label{fig:rho-omega}
\end{figure}

In the examination of $\pi^+ \pi^-$ scattering, we employed a formalism that accounted for contributions from the $Z'$ by modifying the the $\rho$ propagator with insertions of $\rho-Z'$ mixing. In this section, we discuss the way in which one may model the $\omega$-induced feature in $e^+ e^- \to \pi^+ \pi^-$ in this formalism. We start by making the assumption that the dominant contribution due to $\omega$ mixing will come from diagrams in which the $\omega$ mixes with the photon, and subsequently the higher order insertions arise from mixing with the $\rho$. We show example diagrams in Fig.~\ref{fig:rho-omega}. The diagrams in which the $\omega$ does not mix with the $\gamma$ require an additional insertion of mixing at leading order, and we thus take the diagrams with $\gamma-\omega$ mixing as the dominant contribution. 

\begin{figure}[h!]
    \centering
    \includegraphics[width=0.7\textwidth]{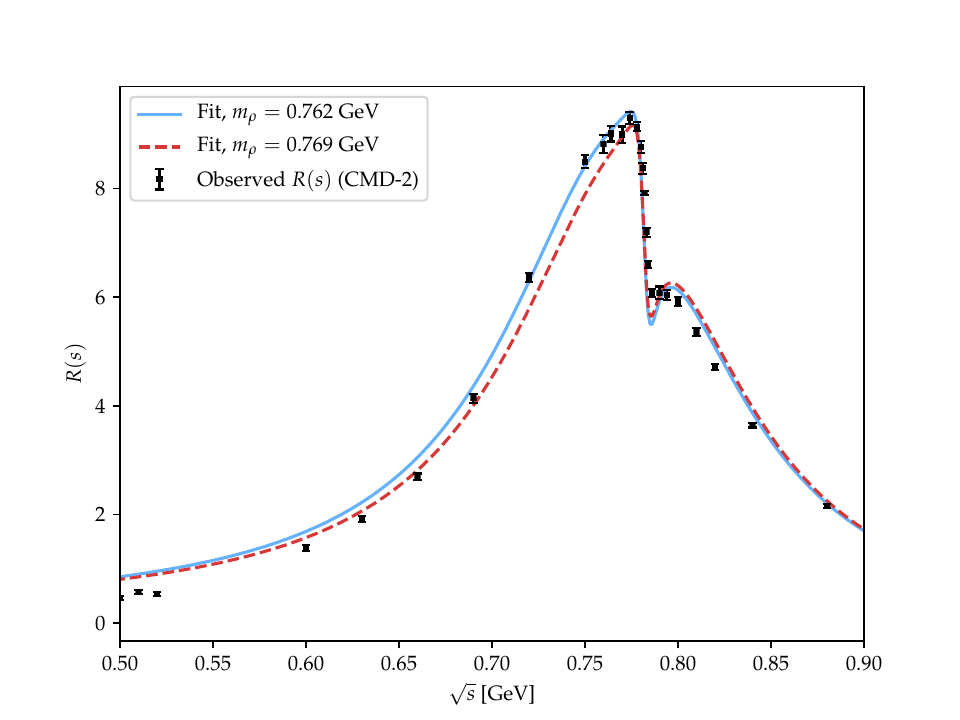}
    \caption{Comparison of the predictions for $R(s)$ from our formalism for $m_{\rho}=0.769$ GeV (red dashed) and $m_{\rho}=0.765$ GeV (blue) compared with the observed $R(s)$ from CMD-2 data (black points).}
    \label{fig:omega_Rhad}
\end{figure}

The expression for the propagator thus includes a new term proportional to the $\omega$ propagator, leading to
\begin{align}
    D(\gamma) \left( \frac{A_{\rho\gamma} + A_{\gamma\omega} D(\omega) A_{\rho\omega}}{1-A_{\rho\gamma}^2 D(\gamma) D(\rho) - A_{\rho\omega}^2 D(\omega) D(\rho)} \right) D(\rho)
\end{align}
where the denominator includes a resummation over high-order insertions of both $\gamma$ and $\omega$. We compare this expression with $R(s)$ as
\begin{align}
    R(s) = \frac{1}{4}\left| D(\gamma) \left( \frac{A_{\rho\gamma} + A_{\gamma\omega} D(\omega) A_{\rho\omega}}{1-A_{\rho\gamma}^2 D(\gamma) D(\rho) - A_{\rho\omega}^2 D(\omega) D(\rho)} \right) D(\rho) \right|^2 \left(\frac{g_{\rho\pi}}{e} \right)^2
\end{align}
where the factor 1/4 arises from the difference in spin of the final state particles. In our analysis, we employ the values $m_{\omega} = 782$ MeV, $\Gamma_{\omega} = 8$ MeV, and $g_{\rho\pi}^2/4\pi=2.8$. We present the results for both the PDG average $m_{\rho}=769$ MeV and for the case in which we allow $m_{\rho}$ to be a fit parameter, finding that the prediction best models the $\rho-\omega$ feature for $m_{\rho}=765$ MeV. For $m_{\rho}=769$ MeV, we find a best fit in the region of the $\rho$ resonance for values around $A_{\rho\gamma}=0.0330$ GeV$^2$, $A_{\gamma\omega}=0.008$ GeV$^2$, and $A_{\rho\omega}=-0.006$ GeV$^2$; for $m_{\rho}=765$ MeV, we find a best fit for $A_{\rho\gamma}=0.0334$ GeV$^2$, $A_{\gamma\omega}=0.009$ GeV$^2$, and $A_{\rho\omega}=-0.006$ GeV$^2$. We show a comparison of our predictions with the data in Fig.~\ref{fig:omega_Rhad}.

\clearpage
\printbibliography
\end{document}